\documentclass[aps,prb,twocolumn,groupedaddress]{revtex4}
\usepackage[bookmarks]{hyperref}
\usepackage[dvips]{graphicx}
\usepackage{amsmath}
\usepackage{amstext}
\usepackage{graphics}
\usepackage{exscale}
\usepackage{epsfig}
\usepackage{changebar}
\usepackage[T1]{fontenc}
\usepackage{bm}
\usepackage{bbm}

\usepackage{version}

\begin{document}
\title[A new renormalization group approach]{A new
 renormalization group approach for
 systems with strong electron correlation} 
\author{K. Edwards and A C Hewson }
\affiliation{${}$Department of Mathematics, Imperial College, London SW7 2AZ,
  UK.}


\date{\today}

\begin{abstract}
The anomalous low energy behaviour observed in metals with 
strong electron correlation, such as in the heavy fermion materials,
is believed to arise  from the scattering of the
itinerant electrons with  low energy spin fluctuations.
In systems with magnetic impurities this scattering leads to the Kondo effect
and a low energy renormalized energy scale, the Kondo temperature
$T_{\rm K}$. It has been generally assumed that these low energy scales
can only be accessed by a non-perturbative approach due to the strength
of the local inter-electron interactions. Here we show that it is  possible to
circumvent this difficulty by  first suppressing the spin fluctuations with  a large magnetic
field.  As a first step field-dependent renormalized parameters are  calculated using 
standard perturbation theory. A renormalized perturbation theory is then used
to calculate the renormalized parameters for a reduced magnetic field strength. 
The process can be repeated and the flow of the renormalized parameters 
continued to zero magnetic field. We illustrate the viability of this approach
for the single impurity Anderson model. The results for 
the renormalized parameters, which  flow as a function of magnetic field,
can be checked with  those from  numerical renormalization group and Bethe ansatz
 calculations.

\end{abstract}
\maketitle

\section{Introduction}

Materials which have been classified as having strong electron correlation
are those which have itinerant electrons in narrow bands arising
from atomic d and f states. The effective on-site inter-electron interaction
for electrons in these shells is of the order, or greater than, the band widths so that the direct
application of perturbation theory to models of these systems,
such as the Hubbard model or periodic Anderson model, is not valid in most
parameter regimes --- particularly in the regimes where anomalous
behaviour is expected.  The same applies for impurity models, such as the
Anderson impurity model, where the anomalous low energy scale, the Kondo
temperature
$T_{\rm K}$, occurs only when the impurity electron is almost
localised and experiences a strong on-site interaction.
For this reason the search has been for non-perturbative techniques
that can handle the strong on-site interactions. Such non-perturbative
techniques as the numerical renormalization group (NRG) \cite{Wil75,KWW80a}
and Bethe ansatz (BA) \cite{AFL83,TW83} 
have had considerable success in tackling most impurity problems,
and some significant progress has been made in  understanding 
the behaviour of lattice models, such as the Mott-Hubbard metal-insulator transition,
using dynamical mean field theory (DMFT) \cite{GKKR96}  in conjunction with a
non-perturbative impurity
solver. Many other aspects of strong correlation behaviour, such as quantum
criticality and the competition between various low temperature broken
symmetry states, both magnetic and superconducting, are still very much open problems where there is a need for
new ideas and techniques.   A feature of strongly correlated systems is that
there are low lying electronic collective excitations, which scatter strongly
the  itinerant electrons. In systems with a strong on-site interaction $U (>0)$,
these are usually spin excitations, arising from the effective ferromagnetic
or antiferromagnetic exchange interactions, which appear explicitly in models
such as the t-J models, which can be derived from a Hubbard model when $U$ is large, where the
antiferromagnetic exchange term
$J\sim t^2/U$ and $t$ is the hopping matrix element.
\par
Here we put to the test a new approach  for accessing the low energy
scales by applying it to an impurity
Anderson model in the strong correlation regime. We have accurate results
for this model from the Bethe ansatz (BA) and the numerical renormalisation group (NRG) which we can
use to check our results. The motivation is to develop a technique which we can
apply to a wider class of models. The Bethe ansatz, though powerful,
is restricted to integrable models, which cover a wide class of
impurity
models, and some one dimensional models such as the Hubbard model, but there
is no generalisation to models of higher dimension. The NRG can only be applied to
impurity models which have  a low degeneracy of impurity states so that the
matrix sizes do not become too large for practical iterative diagonalisation.
In conjunction with the DMFT the NRG  can also be applied to infinite dimensional models which can be mapped into
effective impurity models. 
The density matrix renormalisation group (DMRG) \cite{PWKH98} is an alternative form which has been applied
successfully to one dimensional systems. There are also functional forms of the
renormalisation group (fRG) \cite{KBS10} and non-perturbative RG approaches which are at
present being developed and applied with some success, but the problem of dealing with the strong
correlation of models for systems in two and three dimensions is still an open
challenge.\par
The approach we consider here is based on a renormalised perturbation
theory (RPT). The RPT has been developed and tested in detail for the Anderson
impurity model \cite{Hew93,Hew01}. We begin by giving  a brief description of this application
and then consider the problems in applying the method more generally,
and how they might be overcome.\par
  The Hamiltonian for the Anderson model \cite{And61} is
\begin{eqnarray}
&&H_{\rm AM}=\sum\sb {\sigma}\epsilon\sb {\mathrm{d},\sigma} d\sp {\dagger}\sb
{\sigma}  
d\sp {}\sb {\sigma}+Un\sb {\mathrm{d},\uparrow}n\sb {\mathrm{d},\downarrow} \label{ham}\\
&& +\sum\sb {{ k},\sigma}( V\sb { k,\sigma}d\sp {\dagger}\sb {\sigma}
c\sp {}\sb {{ k},\sigma}+ V\sb { k,\sigma}\sp *c\sp {\dagger}\sb {{
k},\sigma}d\sp {}\sb {\sigma})+\sum\sb {{
k},\sigma}\epsilon\sb {{ k},\sigma}c\sp {\dagger}\sb {{ k},\sigma}
c\sp {}\sb {{
k},\sigma}, \nonumber
\end{eqnarray}
where $\epsilon_{\mathrm{d},\sigma}=\epsilon_{\rm d}-\sigma g\mu_{\rm B} H/2$
is the energy of the localised  level at an impurity site   in a magnetic
field $H$, $U$ the interaction at this local site, and $V_{k,\sigma}$ the
hybridization matrix element to a band of conduction electrons of spin
$\sigma$ with energy $\epsilon_{k,\sigma}=\epsilon_{k}-\sigma g_c\mu_{\rm B} H/2$, where
$g_c$ is the g-factor for the conduction electrons. When $U=0$ the local level
broadens into a resonance, corresponding to a localised quasi-bound state,
whose width depends on the quantity $ \Delta_\sigma(\omega)=\pi\sum\sb {k}|
V\sb {k,\sigma}|\sp 2\delta(\omega -\epsilon\sb { k,\sigma})$. For the
impurity model, where we are interested in universal features, it is usual to
take   a wide conduction band with a flat density of states so that
$\Delta_\sigma(\omega)$ becomes independent of $\omega$,
and can be taken as a constant $\Delta_\sigma$. In this wide band
limit $\Delta_\sigma(\omega)$ will be independent of the magnetic field
on the conduction electrons, so we can effectively put $g_c=0$.
When this is the case $\Delta_\sigma$ is usually taken to be a constant 
$\Delta$ independent of $\sigma$. We also introduce the notation $h=g\mu_{\rm B}H/2$.
\par
In the renormalized perturbation theory approach\cite{Hew93,Hew01} we cast the corresponding
Lagrangian for this model ${\cal L}_{\rm
AM}(\epsilon_{\mathrm{d},\sigma},\Delta,U)$  into  the form,
\begin{equation}
{\cal L}_{\rm AM}(\epsilon_{\mathrm{d},\sigma},
\Delta,U)={ \cal L}_{\rm AM}(\tilde\epsilon_{\mathrm{d},\sigma},
\tilde\Delta_\sigma,\tilde U)+ {\cal L}_{\rm
  ct}(\lambda_{1,\sigma},\lambda_{2,\sigma},\lambda_3),\label{lag}
\end{equation}
where the renormalized parameters,   $\tilde\epsilon_{\mathrm{d},\sigma}$ and 
$\tilde\Delta_{\sigma}$, are defined in terms of the self-energy
$\Sigma_{\sigma}(\omega,h)$ of the one-electron Green's function for the impurity state,
\begin{equation}
G_{\sigma}(\omega,h)={1\over
    \omega-\epsilon_{\mathrm{d}\sigma}+i\Delta-\Sigma_\sigma(\omega,h)},
\label{gf}
\end{equation}
and are given by 
\begin{equation}
\tilde\epsilon_{\mathrm{d},{\sigma}}=z_{\sigma}(\epsilon_{\mathrm{d},{\sigma}} 
+\Sigma_\sigma(0,h)),\quad\tilde\Delta_{\sigma} =z_\sigma\Delta,
\label{ren1}
\end{equation} 
where $z_{\sigma}$ is given by
$z_{\sigma}={1/{(1-\Sigma_{\sigma}'(0,h))}}$.
The renormalized or quasiparticle interaction  $\tilde U(h)$, is defined in terms
of the local total 4-vertex
$\Gamma^{(4)}_{\uparrow\downarrow}(\omega_1,\omega_2,\omega_3,\omega_4; h)$ at zero frequency,
  \begin{equation} 
\tilde U(h)=z_{\uparrow}z_{\downarrow}\Gamma^{(4)}_{\uparrow\downarrow}(0,0,0,0; h).
\label{ren2}\end{equation}
It will be convenient to rewrite the spin dependent quasiparticle energies
in the form,
$\tilde\epsilon_{{\rm d},\sigma}=\tilde\epsilon_{\rm d}(h)-\sigma h\tilde\eta(h)$,
where
\begin{equation}
\tilde\epsilon_{\mathrm{d}}(h)={1\over
  2}\sum_\sigma\tilde\epsilon_{\mathrm{d},{\sigma}},\quad 
\tilde\eta(h)={1\over
  2h}\sum_\sigma \sigma\tilde\epsilon_{\mathrm{d},{\sigma}},
\end{equation}
where $\tilde\epsilon_{\rm d}(h)$ and $\tilde\eta(h)$ are both  even functions of the magnetic field
$h$. 
\par
In terms of the renormalised parameters, the Green's function takes the form,
\begin{equation}
G_{\sigma}(\omega,h)={z_\sigma(h)\over
    \omega-\tilde\epsilon_{\mathrm{d}}(h)+\sigma\tilde h+i\tilde\Delta_\sigma(h)-\tilde\Sigma_\sigma(\omega,h)},
\label{ngf}
\end{equation}
where $\tilde h=\tilde\eta(h)h$, and $\tilde\Sigma_\sigma(\omega,h)$ is the
renormalised self-energy given by 
\begin{equation}
\tilde\Sigma_{\sigma}(\omega,h)=z_\sigma(\Sigma_{\sigma}(\omega,h)-
\Sigma_{\sigma}(0,h)-\omega\Sigma'_{\sigma}(0,h)).
\label{rse}
\end{equation}
 The propagator in the RPT is the free quasiparticle Green's function,
 \begin{equation}
\tilde G_{\sigma}(\omega,h)={1\over
    \omega-\tilde\epsilon_{\mathrm{d}}(h)+\sigma\tilde h+i\tilde\Delta_\sigma(h)}.
\label{qpgf}
\end{equation}
The expansion is carried out in  powers of the  
interaction $\tilde U(h)$ for the complete Lagrangian defined in equation
(\ref{lag}). 
 The counter term part of the Lagrangian $ {\cal L}_{\rm
  ct}(\lambda_{1,\sigma},\lambda_{2,\sigma},\lambda_3)$, given by
\begin{eqnarray}
&&{\cal L}_{\rm
  ct}(\lambda_{1,\sigma},\lambda_{2,\sigma},\lambda_3)=\\
\quad &&\sum_\sigma\bar
  d_\sigma(\tau)(\lambda_{2,\sigma}\omega+\lambda_{1,\sigma})d_\sigma(\tau)+\lambda_3n_{d,\uparrow}(\tau)
n_{d,\downarrow}(\tau),\nonumber
\end{eqnarray}
where $\bar d_\sigma(\tau)d_\sigma(\tau)=n_{d,\sigma}(\tau)$, and
$\bar d_\sigma(\tau)$ and $ d_\sigma(\tau)$ are the Grassmann fields
corresponding to the impurity creation and annihilation operators,
which are integrated over in the  calculation of the partition function.  This
part of the Lagrangian essentially takes care of any 
overcounting.
The parameters, $\tilde\epsilon_{\mathrm{d},{\sigma}}$, $\tilde\Delta_\sigma$
 and $\tilde U$, have been taken to be the fully renormalized ones, and the
 counter term parameters, $\lambda_{1,\sigma}$, $\lambda_{2,\sigma}$ and $\lambda_3$, are
 required to cancel any further renormalisation. This leads to the
 renormalisation or over-counting conditions,
\begin{equation}
\tilde\Sigma_\sigma(0,h)=0,\quad {\partial\tilde\Sigma_\sigma(\omega,h)
\over \partial\omega}\Big|_{\omega=0}=0,
\label{rc1}
\end{equation}
and 
\begin{equation}
\tilde\Gamma_{\uparrow,\downarrow}(0,0,0,0;h)=\tilde U(h).
\label{rc2}
\end{equation}
The counter terms  are completely
determined by these conditions.\par
{\em} Exact results for the Fermi liquid regime, which were first derived in a
 phenomenological approach by Nozi\`eres \cite{Noz74}, can be derived from the RPT,
working only to second order in $\tilde U$.
 We give some of these results
for the particle-hole symmetric model. In this case, $\tilde\epsilon_d(h)=0$
and $\tilde\Delta_\sigma$ is independent of $\sigma$, so we drop the $\sigma$
index for this quantity. The free quasiparticle density of states
$\tilde\rho_\sigma(\omega,h)$, given by
the spectral density of the free quasiparticle Green's function in equation
(\ref{qpgf}), takes the form,
\begin{equation}
\tilde\rho_\sigma(\omega,h)=\rho^{(0)}_{\sigma}(\omega,\tilde h,\tilde\Delta(h)),
\label{qpdos}
\end{equation}
where $\rho^{(0)}_{\sigma}(\omega, h,\Delta)$ is the local density of states
for the non-interacting system,
\begin{equation}
\rho^{(0)}_{\sigma}(\omega, h,\Delta)={1\over\pi}{\Delta\over (\omega+\sigma h)^2+\Delta^2}.
\label{qpdos2}
\end{equation}
As $\rho^{(0)}_{\sigma}(\omega, h,\Delta)$ becomes independent of $\sigma$ for
$\omega=0$, we can  drop the $\sigma$ index when $\omega=0$.\par

The  impurity contribution to the coefficient of the electronic
specific
 heat $\gamma(h)$ is directly proportional to the quasiparticle
density of states,
\begin{equation}
\gamma(h)={2\pi^2k^2_{\rm B}
\over 3}\tilde\rho(0,h),
\label{gamqp}
\end{equation}
where $k_{\rm B}$ is the Boltzmann constant.
The induced magnetisation $M(h)$ is given by 
 $M(h)=g\mu_{\rm B}m(h)$, where 
\begin{equation}
m(h)={1\over 2}(\langle n_{{\rm d}\uparrow}\rangle -\langle n_{{\rm
    d}\downarrow}\rangle )=m_0(\tilde h,\tilde\Delta(h)),
\label{magqp}
\end{equation}
where $\langle n_{{\rm d}\sigma}\rangle$ is the expectation value of the
occupation number of the impurity site, 
$n_{{\rm d}\sigma}$, and $m_0(h,\Delta)$ is the magnetisation
for the non-interacting model given by
\begin{equation}
m_0(h,\Delta)={1\over\pi}\,{\rm tan}^{-1}
\left({ h\over\Delta}\right).
\label{magqp2}
\end{equation}
 The 
longitudinal  susceptibility  $\chi_l(h)$ (in units of $(g\mu_{\rm B})^2$)
is given by
\begin{equation}
\chi_l(h)={\tilde\rho(0,h)(1+\tilde
U(h)\tilde\rho(0,h))/2},
\label{chil}
\end{equation}
and the local charge susceptibility  $\chi_c(h)$ is given by
\begin{equation}
\chi_c(h)=2{\tilde\rho(0,h)( 1-\tilde
U(h)\tilde\rho(0,h))}.
\label{chic}
\end{equation}
The  perpendicular susceptibility $\chi_\perp(h)$  is given by 
\begin{equation}
\chi_\perp(h)=\frac{m(h)}{2h},
\label{chit}
\end{equation}
in the limit of zero transverse field \cite{Hew06}.

Asymptotically exact results can also be derived  for 
low energy dynamic spin and charge susceptibilities by taking into
account repeated quasiparticle scattering \cite{Hew06}. The longitudinal
dynamic spin susceptibility is given by 
\begin{equation}
\chi_l(\omega,h)={\Pi_l(\omega,\tilde h,\tilde\Delta(h))\over
                   2(  1-\tilde U_s(h)\Pi_l(\omega,\tilde h,\tilde\Delta(h)))},
\label{chilw}
\end{equation}
for the symmetric model,
where 

   $$\Pi_{l}(\omega,h,\Delta)={\Delta\over{\pi( h^2+\Delta^2)}}\quad{\rm for}\quad\omega= 0,$$
\begin{equation}={-\Delta\over{\pi\omega(\omega+2i\Delta)}}\left\{{\rm
      ln}\left({{\omega+i\Delta- h}\over{i\Delta-h}}\right)+
{\rm ln}\left({{\omega+i\Delta+ h}\over{i\Delta+ h}}\right)\right\},\end{equation}                
 for $\omega\ne 0$, where $\tilde U_s(h)$ is the effective quasiparticle interaction
in the longitudinal channel given by $\tilde U_s(h)=\tilde U(h)/(1+\tilde
U(h)\tilde\rho(0,h))$.

The corresponding expression for the transverse susceptibility
$\chi^{+-}(\omega,h)$ which we denote by $\chi_t(\omega,h)$ is
\begin{equation}
\chi_t(\omega,h)={\Pi_t(\omega,\tilde h,\tilde\Delta(h))\over
                     1-\tilde U_t(h)\Pi_t(\omega,\tilde h,\tilde\Delta(h))}.
\label{chitw}
\end{equation}
where
$$\Pi_t(\omega,h,\Delta)={i\over{\pi(i\Delta- h)}} -
{1\over 2\pi \Delta}{\rm
  ln}\left({{i\Delta-h}\over{i\Delta+
  h}}\right)\quad\omega=
  -2 h,$$
\begin{eqnarray}
&&={-i\over\pi}\left\{{1\over \omega+2
  h+2i\Delta}{\rm ln}\left({\omega+i\Delta+
    h\over i\Delta+h}\right)\right.-\nonumber\\
&&\quad\quad{1\over \omega+2 h}
\left.{\rm ln}\left({\omega+i\Delta+ h\over i\Delta-
    h}\right)\right\}\quad\omega\ne -2h.
\end{eqnarray}              

The effective quasiparticle interaction in this channel $\tilde U_t(h)$ 
is determined by the condition that the dynamic transverse susceptibility at
$\omega=0$ it is equal to twice the static
perpendicular susceptibility given in equation (\ref{chit}),
$\chi_t(0,h)=2\chi_\perp(h)$.
This gives 
\begin{equation}
\chi_t(0,h)={{m_0(\tilde h,\tilde\Delta)\over \tilde h} \over
                     1-{\tilde U_t(h)m_0(\tilde h,\tilde\Delta)\over \tilde h}}=
{m_0(\tilde h,\tilde\Delta)\over h},\label{static_chit}
\end{equation} 
on using the result $\Pi_t(0, h,\Delta)=m_0(h,\Delta)/ h$ for $\omega=0$.
We then find $\tilde U_t(h)$ must be such that
\begin{equation}
\tilde h=h+\tilde U_t(h)m(h).
\label{rptmf}
\end{equation}
With  $\tilde U_t(h)$ defined as in equation (\ref{static_chit}), the {\it
  exact} expression for the  magnetisation takes the form of a mean field equation,
\begin{equation}
m(h)= {1\over \pi}{\rm tan}^{-1}\left ({h+\tilde
    U_t(h)m(h)\over\tilde\Delta(h)}\right).
\label{rptmf2}
\end{equation}

The only equations that we have quoted that are not exact are the equations
for the dynamic susceptibilities $\chi_l(\omega,h)$ and $\chi_t(\omega,h)$
for finite frequency $\omega\ne 0$. They have nevertheless been shown to
provide a very good approximation to the NRG results for these quantities
over the whole low frequency range and satisfy the Korringa-Shiba relation
exactly (for more details see reference \cite{Hew06}).\par

To evaluate these formulae we need to know the renormalised parameters.
In the Kondo regime, in the absence of a magnetic field, these can be reduced to a single parameter \cite{Hew93}, the Kondo temperature
$T_{\rm K}$ ($\chi_l(0)=1/4T_{\rm K}$) such that 
\begin{equation}
\tilde U(0)=\pi\tilde\Delta(0)=4T_{\rm K},
\label{KT}
\end{equation}
 but this parameter  is still undetermined. The most accurate way
of
calculating the parameters in terms of the 'bare' parameters of the model,
$\epsilon_d$, $\Delta$ and $U$, is
an indirect one, from an analysis of the low energy fixed point in an NRG 
calculation.  The procedure for doing this is described elsewhere \cite{HOM04} . This
approach works very well but restricts the method to 
models where we can apply the NRG, which means one  where we already have  a solution. 
It is, nevertheless,  a useful adjunct to the NRG, enabling one to calculate many quantities
more easily and more accurately. However, we want to develop a way of
calculating the renormalised parameters which is independent of the NRG and
can be applied more widely. The obvious method would be to calculate them
directly from their definitions in equations (\ref{ren1}) and (\ref{ren2}),  but this would again seem  to require having a
solution, at least for the low energy behaviour of the self-energy. A direct
perturbation approach to calculate the self-energy would not seem
 to offer the possibility of accessing the strong correlation regime, as we know
for the symmetric model, low order perturbation theory is unreliable for
$U/\pi\Delta>1$,  and no-one has so far succeeded in summing a subclass of terms
to give the correct low energy behaviour.\par
\begin{figure}[!htbp]
   \begin{center}
     \includegraphics[width=0.45\textwidth]{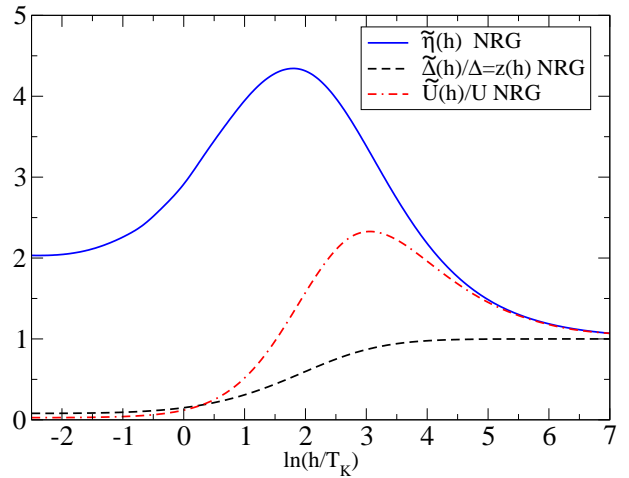}
 \caption{Plots of the renormalised parameters,
  $\tilde\Delta(h)/\Delta$, 
  $\tilde U(h)/U$ and  $\tilde\eta(h)$, for
 the symmetric Anderson model, calculated using the NRG for the case $\pi\Delta=0.1$, $U/\pi\Delta=3$  as a function of the logarithm of the magnetic
 field $h/T_{\rm K}$, where $T_{\rm K}=\pi\tilde\Delta(0)/4=0.002$.}
     \label{rp1}
   \end{center}
 \end{figure}

 We do know, however, that for this
model the large renormalisation effects arise for large positive $U$ from the
scattering of local low energy spin fluctuations. If we apply a strong local
magnetic field these spin fluctuations are suppressed, so the renormalisation
effects are much weaker, and perturbation theory can be applied in this case, even in the
regime $U/\pi\Delta\gg 1$ \cite{HZ84}. This implies that in the limit of a very large
magnetic field, we can evaluate the renormalised parameters directly from
equations (\ref{ren1}) and (\ref{ren2}), using perturbation results for the self-energy. This
assumption can be checked, as we have results for the Anderson model of the
renormalised parameters for any value of the magnetic field, deduced using
the NRG \cite{Hew05,HBK06,BH07a}. An example is shown in figure \ref{rp1}  for the symmetric Anderson model,
with a value of $U/\pi\Delta=3$, $\pi\Delta=0.1$, such that in very low magnetic fields we are
in the strong correlation regime with a Kondo temperature $T_{\rm K}=0.002$. The degree of
renormalisation in zero field can be estimated from $1/z=\Delta/\tilde\Delta$,
corresponding to a mass enhancement factor, which in this case gives $1/z\sim
12.7$.
  \vspace*{0.7cm}
 We can see from figure \ref{rp1}  that 
$z$ increases monotonically with increase in the value of the  magnetic field.
It does, however, need extremely large field values, such that $h\sim U$,
before the renormalisation effects are completely suppressed. The other two
parameters, $\tilde\eta(h)$ and $\tilde U(h)$, do not simply increase monotonically
as the magnetic field is increased. Initially they increase rather slowly, and
then rise much more rapidly to a peak value, and then in the extreme large
field limit approach their bare values.  The initial increase is not
surprising because the spin fluctuations are being suppressed and the
quasiparticles are becoming less renormalised, the first stage of their
undressing.  We see that the peak value of $\tilde U(h)$ is greater
than the bare value $U$. This is because at this point we are approaching
the regime where mean field theory and RPA are applicable. In this regime,
RPA corresponds to substituting $\tilde U_s(h)\to U$ and $\tilde U_t(h)\to U$,
with $\tilde\Delta(h)\to\Delta$ into equations (\ref{chilw}) and (\ref{chitw}),
giving 
 $\tilde h\to \bar h =h+Um(h)$ and from equation (\ref{chil}), $\tilde U(h)\to U/(1-U\rho^{(0)}(0,\bar h,\Delta))$.
  This can  explain why $\tilde U(h)$ can be
 enhanced over the bare value $U$.
As
 $\rho^{(0)}(0,\bar h,\Delta)\to 0$
 as $h\to\infty$,
the enhancement disappears for very large field values.
The other term
contributing to $\tilde U(h)$ from  the $z$-factor in (\ref{ren2}) has the opposite effect.
In the mean field regime for very large fields, where $z(h)=1$, it
has no effect but  plays a dominant role in reducing    
$\tilde U(h)$ in the weak field regime where $z(h)$ is small.
\par
 The peak in $\tilde\eta(h)$ can be
explained in a similar way. For large fields we have  the mean field result
for $\tilde\eta(h)=1+Um(h)/h$, so that $\tilde\eta(h)\to 1$ as $h\to\infty$
because $m(h)\to 1/2$. It follows from equations
(\ref{magqp}) and (\ref{chil}) that in the limit $h\to 0$  the value of  $\tilde\eta(0)$ is equal
to the Wilson $\chi/\gamma$ ratio, which in the strongly correlated (Kondo)
regime takes the value 2.  The fact the $z(h)$ increases  as $h^2$  for small $h$ explains
the initial increase of  $\tilde\eta(h)$ with $h$. That a  peak should
occur  at an intermediate
field value follows as the  likely behaviour from the extrapolated trends
at low and large field values.
\par
We have conjectured that we should be able to explain the NRG results for the
renormalised parameters in the  large magnetic field limit using the leading
perturbational corrections arising from mean field theory and RPA. The
question then arises: Can we find a way to continue the process to
lower magnetic fields, as the renormalised parameters in large
fields are continuously connected to the strongly renormalised values
in the weak field regime? We know that if we start at the other limit,
with the zero field renormalised parameters known, we can use the
RPT to calculate the renormalised self-energy in a weak field $h$ \cite{Hew01}.
From this result, we could then calculate the change in the renormalised parameters
due to the introduction of the weak magnetic field.  We could generalise this
idea by considering that we know the parameters for an  arbitrary magnetic
field value $h$, and use these to calculate the self-energy for a system
with a slightly smaller or larger magnetic field $h\pm\delta h$, and hence
deduce the renormalised parameters for these neighbouring magnetic field
values.  This would give a set of scaling equations for the renormalised
parameters as a function of the magnetic field. If we can obtain good starting
values from perturbation theory in the large field limit, we should be able
 to  reduce iteratively the magnetic field,
and scale into the strongly correlated weak field regime. The aim of this
paper is to test the feasibility of this scheme, and
 we use the results from the NRG and Bethe ansatz to test any approximation. We begin first of all looking at the behaviour in the limit of
large magnetic field.
   \section{Large magnetic field limit}

A formally exact expression can be written down for the self-energy
$\Sigma_\sigma(\omega)$ in terms of skeleton diagrams, where the propagators
correspond to the full many-body Green's function $G_{d,\sigma}(\omega)$, as given in equation
(\ref{gf}), and the full vertex
$\Gamma^{(4)}_{\sigma,-\sigma}(\omega_1,\omega_2,\omega_3,\omega_4)$,
$$
\Sigma_\sigma(\omega)= U\langle n_{d,-\sigma}\rangle+U\int\int G_{d,\sigma}(\omega+\omega')
G_{d,-\sigma}(\omega'') $$
\begin{equation}    
   G_{d,-\sigma}(\omega''-\omega')
\Gamma^{(4)}_{\sigma,-\sigma}(\omega+\omega',\omega''-\omega',\omega'',\omega){d\omega'\over
    2\pi i}{d\omega''\over
    2\pi i}
\label{exactsee}
\end{equation}

\begin{figure}[!htbp]
   \begin{center}
     \includegraphics[width=0.45\textwidth]{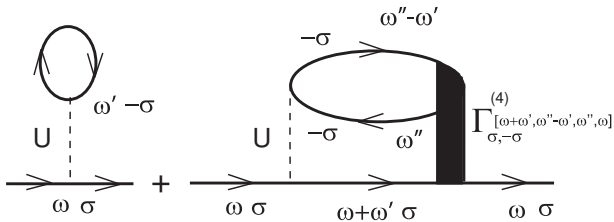}
 \caption{The skeleton diagrams for the self-energy $\Sigma_\sigma(\omega)$
 in terms of the 4-vertex $\Gamma_{\sigma,-\sigma}^{(4)}$}
     \label{exactsed}
   \end{center}
 \end{figure}
 \noindent
A diagrammatic representation of this equation is given in figure
\ref{exactsed}. The skeleton diagrammatic formulation of the perturbation
theory is particularly useful when self-consistent approximations are used.
The simplest self-consistent approach is the mean field theory where only the
tadpole diagram, corresponding to the first term in equation 
(\ref{exactsee}), is taken into account. For the particle-hole symmetric model
this gives  $\Sigma_\sigma(\omega)=-\sigma Um(h)$, where $m(h)$
is determined self-consistently from the equation,
\begin{equation}
m(h)=
{1\over\pi}\,{\rm tan}^{-1}
\left({\bar h\over\Delta}\right),
\label{magmf}
\end{equation}
where $\bar h=h+Um(h)$. Note that we have not included the constant term $U/2$ in the self-energy
because for the symmetric model it can always be absorbed into the energy level $\epsilon_d$ ($=-U/2$)
to give $\epsilon_d=0$.
 From equation (\ref{rptmf}) this  implies that in the
mean field regime
$\tilde U_t(h)\to U$.
 For this approximation
we
deduce  the renormalised parameters, $\tilde\eta(h)$ and $\tilde\Delta(h)$.
As the self-energy is independent of $\omega$, it follows that $z(h)=1$, so
\begin{equation}
\tilde\eta(h)=1+{U\over \pi h}\,{\rm tan}^{-1}
\left({\bar h\over\Delta}\right), \quad \tilde\Delta(h)=\Delta,
\end{equation}
where $m(h)$ is determined self-consistently from (\ref{magmf}).
We can deduce the static longitudinal susceptibility by differentiating
 $m(h)$ with respect to $h$, and from equation (\ref{chil}) deduce
$\tilde U$. The result is $\tilde U_s(h)=U$, or equivalently,
\begin{equation}
\tilde U(h)={U\over 1-U\rho^{(0)}(0,\bar h,\Delta)}.
\end{equation}
 We anticipated this result in the previous section, based
on the RPA approximation for the longitudinal dynamical  susceptibility.\par
\begin{figure}[!htbp]
   \begin{center}
     \includegraphics[width=0.45\textwidth]{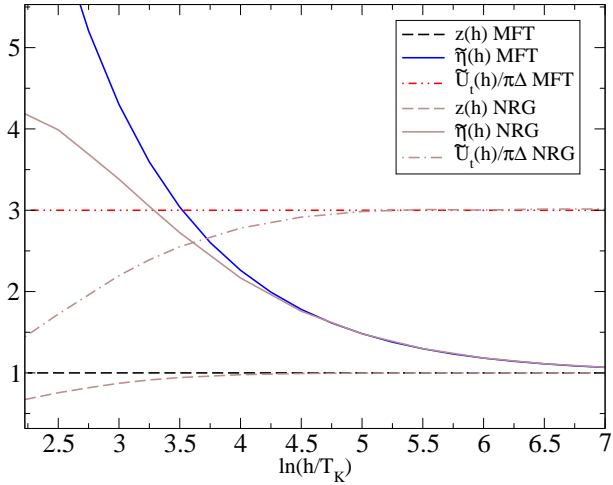}
 \caption{A comparison  of the renormalised parameters,  $\tilde\eta(h)$
  $z(h)=\tilde\Delta(h)/\Delta$   and 
  $\tilde U_t(h)/U$,  from mean field theory
  with those calculated using the NRG, for 
 the symmetric Anderson model, with $\pi\Delta=0.1$, $U/\pi\Delta=3$  as a function of the logarithm of the magnetic
 field $h/T_{\rm K}$, where $T_{\rm K}=\pi\tilde\Delta(0)/4$.}
     \label{MFrp}
   \end{center}
 \end{figure}

We now have a 1-1 correspondence between the RPT results and
the mean field/RPA equations in the transverse spin scattering channel,
and we conjecture that asymptotically for very large magnetic field values,
\begin{equation}
\tilde h\to \bar h=h+Um(h),\quad \tilde\Delta(h)\to \Delta,\quad \tilde U_t(h)\to U.
\end{equation}
In figure \ref{MFrp} we compare the mean field results 
for  $z(h)=\tilde\Delta(h)/\Delta$,   
  $\tilde U_t(h)/U$ and  $\tilde\eta(h)$, with the corresponding results
  calculated using the NRG as a function of ${\rm ln}(h/T_{\rm K})$, for the model for  $\pi\Delta=0.1$
and  $U/\pi\Delta=3$. As expected the mean field results are in agreement with
the NRG results for asymptotically large magnetic fields. The mean field
results give a good approximation for these parameters for values of
$h>90T_{\rm K}$ (${\rm ln}(h/T_{\rm K})> 4.5)$. 
Though the mean field results are only valid for very large magnetic field
values,
we can build upon them by expanding about the mean field solution. The propagators in the perturbation
expansion 
now take into account the mean field self-energy  so that the tadpole
diagrams, or mean field insertions,  no
longer appear explicitly.  \par
 \vspace*{0.7cm}
\begin{figure}[!htbp]
   \begin{center}
     \includegraphics[width=0.45\textwidth]{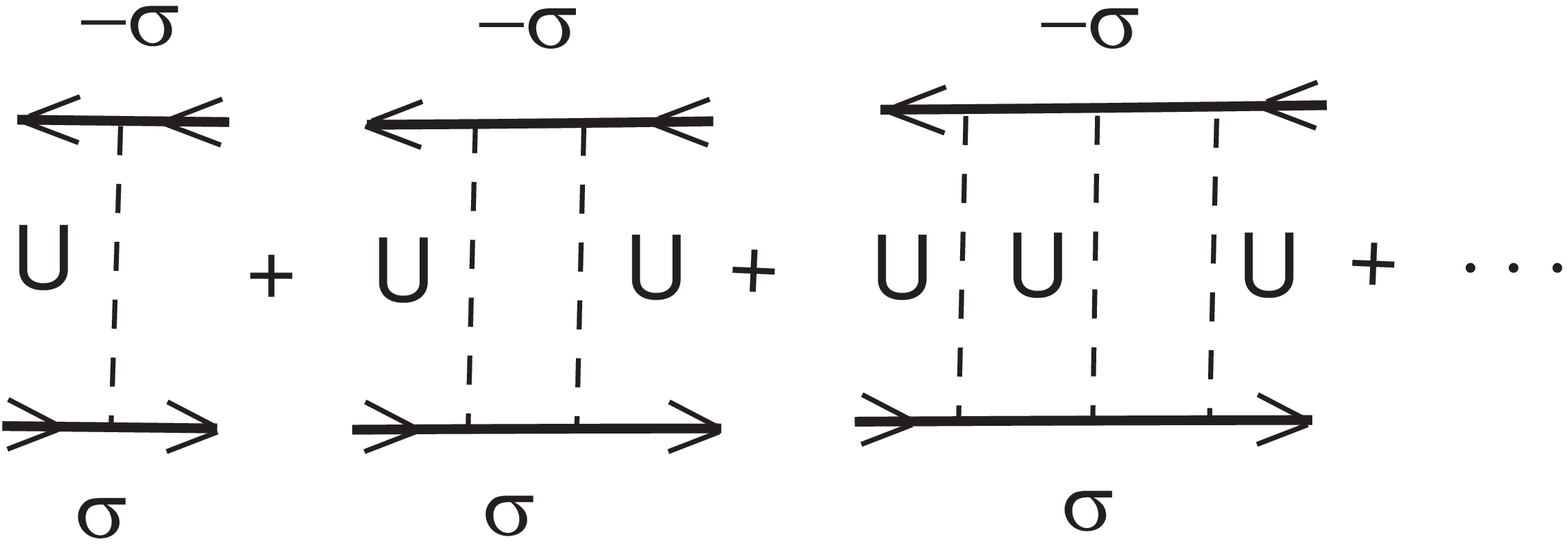}
 \caption{ The RPA diagrams that contribute to the 4-vertex $\Gamma^{(4)}_{\sigma,-\sigma}(\omega+\omega',\omega''-\omega',\omega'',\omega)$ }
     \label{gammarpa}
   \end{center}
 \end{figure}

 \noindent
 For $U>\pi\Delta$ in the absence of a magnetic field,
the mean field solution predicts a state with a local magnetic moment, such
that it 
costs no energy to flip the local moment. This is reflected in the dynamic
transverse spin susceptibility, calculated with mean field propagators, which
develops a singularity at $\omega=0$.
  Hence, the most important corrections to the self-energy 
are likely to arise from these spin flip scattering processes. In evaluating
the self-energy using equation (\ref{exactsee}) we need to include
the dominant spin flip scattering terms contributing to the 4-vertex
 $\Gamma^{(4)}_{\sigma,-\sigma}(\omega+\omega',\omega''-\omega',\omega'',\omega)$, which are illustrated in figure \ref{gammarpa}. These diagrams are the same as  those which are taken into account
in the RPA expression for the transverse dynamic susceptibility
$\chi_t(\omega)$. Summing this class of diagrams
gives as an approximation for
the 4-vertex,
$\Gamma^{(4)}_{\uparrow,\downarrow}(\omega+\omega',\omega''-\omega',\omega'',\omega)$,
\begin{equation}
{U\over 1-U\Pi_t(\omega',\bar h,\Delta)}
\label{RPAgamma4}
\end{equation}
Substituting this result into equation (\ref{exactsee}) gives the result,
\begin{equation}
\Sigma_{\uparrow}(\omega,h)=U^2\int
G^{\rm mf}_{\downarrow}(\omega+\omega',\bar h)\chi_t(\omega',h){d\omega'\over 2\pi i},
\label{RPAse}
\end{equation}
where the Green's function $G^{\rm mf}_{\downarrow}(\omega,\bar h)$ is the mean field
propagator and the
transverse dynamic susceptibility $\chi_t(\omega,h)$ is calculated in
the RPA. There is a similar expression for the spin down self-energy,
but with particle-hole symmetry they can be reduced to a single  equation as
 $G_{\downarrow}(\omega,h)=-G_{\uparrow}(-\omega,h)$.\par
\begin{figure}[!htbp]
   \begin{center}
     \includegraphics[width=0.45\textwidth]{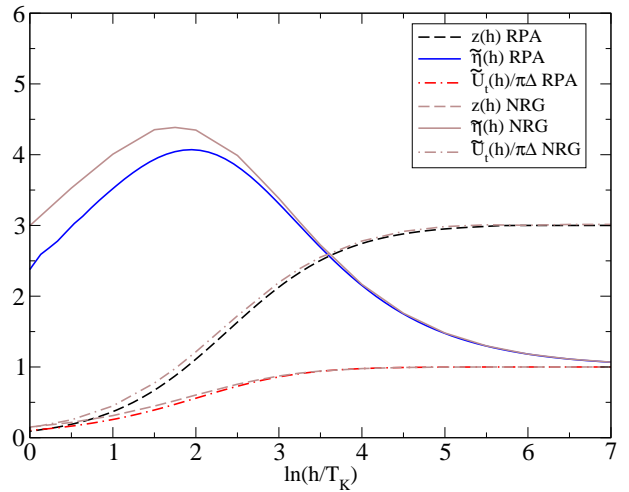}
 \caption{A comparison  of the renormalised parameters,  $\tilde\eta(h)$
  $z(h)=\tilde\Delta(h)/\Delta$   and 
  $\tilde U_t(h)/U$, using the RPA propagator in
  calculating the self-energy
  with those calculated using the NRG, for 
 the symmetric Anderson model, with $\pi\Delta=0.1$, $U/\pi\Delta=3$  as a function of the logarithm of the magnetic
 field $h/T_{\rm K}$, where $T_{\rm K}=\pi\tilde\Delta(0)/4$.}
     \label{rpa_rp}
   \end{center}
 \end{figure}

 \noindent
As the self-energy acquires a dependence on the
frequency $\omega$ in this approximation, in the calculation of the renormalised
parameters, there will be a change of  the quasiparticle weight factor $z(h)$
from the mean field value 1. We can see from the results for the renormalised
parameters calculated using this result, shown in figure \ref{rpa_rp}, that this leads
to a significant improvement in comparing the results with those calculated
using the NRG. The  renormalised parameters are plotted as a function of
${\rm ln}(h/T_{\rm K})$  for the same parameter
set as in figure \ref{MFrp}, $\pi\Delta=0.1$, $U/\pi\Delta=3$.
The values for both  $z(h)$ and $\tilde U_t(h)$ are now seen to decrease
with a decrease in $h$ in a rather similar way to the NRG results. The value
of $\tilde\eta(h)$ also develops a peak as the magnetic field value is reduced
in a similar way to the NRG results, but the calculations for this quantity 
break down for  ${\rm ln}(h/T_{\rm K})<-0.5$.
 The results now constitute a good
approximation to the NRG results down to a magnetic value $h\sim 27 T_{\rm K}$
corresponding to ${\rm ln}(h/T_{\rm K})\sim 3.4$. 
 \par
\vspace*{0.7cm}
 \noindent

We have derived, using perturbation theory,  asymptotically exact expressions
for the renormalised parameters for very large magnetic field values.
 However, the magnetic field regime where these results are valid is completely
outside the range that can be realised experimentally, except possibly for
some quantum dot systems. We need  a way of extending the
calculations to much lower magnetic field values. It should be possible, in
principle at least, to improve on these results by systematically taking
corrections  to the RPA into account.  However, this gets more and
more
difficult, involving multiple frequency integrations and more complex
self-consistent equations to solve. Our aim here is a  limited one, the calculation
of the renormalised parameters which  depend only on the form of the
self-energy in the low frequency regime,  and as a result we can adopt a different
strategy for carrying out higher order calculations.
We  return to the idea discussed at the end of the Introduction, that we might
be able to use the RPT, with renormalised parameters for a magnetic field $h$,
to calculate the renormalised self-energy at a reduced fields value $h-\delta
h$, and hence derive the renormalised parameters for the lower magnetic field
value  $h-\delta
h$. 
\section{Calculations using the RPT}
The
RPT approach differs from standard perturbation theory as it includes counter
terms. We should consider
how to handle these terms, and test various  approximations
for calculating the renormalised self-energy $\tilde\Sigma_\sigma(\omega,h)$
in a given magnetic field $h$, using the renormalised parameters, $\tilde\eta(h)$, $\tilde\Delta(h)$ and $\tilde U(h)$,
for the {\em same} magnetic field value $h$, before considering how to use them for a
system with a reduced magnetic field.\par 
 When  RPT calculations
are carried out to a specific order $n$ in powers of $\tilde U$, we can handle the
counter terms by also  expanding them formally in powers of $\tilde U$, and
taking into account all diagrams generated to order $n$, including those
involving the counter terms.
Counter term contributions of order greater then $n$ will not contribute and 
the terms up to order $n$ can be determined, order by order, by requiring them to satisfy the
renormalisation conditions given in equations (\ref{rc1}) and (\ref{rc2})
(see reference \cite{Hew01} for a specific example). We need a more general
procedure, however,  for calculations based on summing subclasses of diagrams taken to infinite order.
\par
 There are five  counter terms to deal with, four of them, $\lambda_{1,\sigma}$
and $\lambda_{2,\sigma}$ involve
one-body terms, and the fifth is an interaction term $\lambda_3$. The
contributions arising from the diagrams  for the counter terms, $\lambda_{1,\sigma}$
and  $\lambda_{2,\sigma}$, alone, can be
taken into account fully as they do not involve the interaction term. The interaction counter term $\lambda_3$ can be added to the interaction
$\tilde U$, and the expansion carried out in powers of the total
interaction term  $\tilde U+\lambda_3$. The value of $\lambda_3$ is then
required to satisfy the renormalisation condition in equation (\ref{rc2}).
The local Green's function subject to a magnetic field $h$,
using the renormalised parameters for the  field $h$,  takes the form, 
\begin{eqnarray}
&&G_{\sigma}(\omega,h)=\label{ngf3}\\
&&{z(h)\over
    \omega+\sigma\tilde h+i\tilde\Delta(h)+\lambda_{1,\sigma}+\omega\lambda_{2,\sigma}-\tilde\Sigma^S_\sigma(\omega,h)}.\nonumber
\end{eqnarray}
where $\tilde\Sigma^S_\sigma(\omega,h)$ is the self-energy  calculated 
using the quasiparticle propagator, $\tilde G_{\sigma}(\omega,h)=G_{\sigma}(\omega,h)/z(h)$,
which includes the two counter terms, $\lambda_{1,\sigma}$
and  $\lambda_{2,\sigma}$. 
This means that
the self-energy $\tilde\Sigma^S_\sigma(\omega,h)$ will be a function of $\lambda_{1,\sigma}$ and $\lambda_{2\sigma}$, as well as
  $\lambda_3$, so that the renormalisation conditions in equations (\ref{rc1})
take the form,
\begin{equation}
 \tilde\Sigma^{S}_\sigma(0,h,\lambda_{1,\sigma'},\lambda_{2,\sigma'},\lambda_3)=\lambda_{1,\sigma},
\label{rc3}
\end{equation} 
 and
\begin{equation}
{\partial\tilde\Sigma^S_\sigma(\omega,h,\lambda_{1,\sigma'},\lambda_{2,\sigma'},\lambda_3)
\over \partial\omega}\Big|_{\omega=0}=\lambda_{2,\sigma}.
\label{rc4}
\end{equation}
Given a result for $\tilde\Sigma^S_\sigma(0,h)$, 
calculated from a particular subset of diagrams, the conditions in equations
(\ref{rc3}) and (\ref{rc4}) generate self-consistent  equations which must be
solved to determine the counter terms, $\lambda_{1,\sigma}$ and $\lambda_{2,\sigma}$.\par
We can now perform calculations equivalent to mean field and the RPA but
in the RPT framework. At the first mean field stage we only take the tadpole
diagrams into account giving $\Sigma^S_\sigma(\omega,h)=-\sigma(\tilde
U+\lambda_3)m(h)$. In this approximation from equations (\ref{rc3}) and
(\ref{rc4}) we get $\lambda_{1,\sigma}=\sigma(\tilde
U+\lambda_3)m(h)$ and $\lambda_{2,\sigma}=0$, so
\begin{equation}G^{\rm mf}_{\sigma}(\omega,h)=
{1\over
    \omega+\sigma\tilde h+i\tilde\Delta(h)}.
\label{mfqpgf}
\end{equation}
 The next stage is to calculate the
transverse susceptibility $\chi_t(\omega,h)$, using propagators which include
the mean field insertions. The $\lambda_{1,\sigma}$ in the propagator given in equation
(\ref{nqpgf}) cancels the mean field term,    the propagator now becomes
the free propagator given in equation (\ref{qpgf}). The result for
$\chi_t(\omega,h)$ is 
\begin{equation}
\chi_t(\omega,h)={\Pi_t(\omega,\tilde h,\tilde\Delta(h))\over
                     1-(\tilde U+\lambda_3)\Pi_t(\omega,\tilde h,\tilde\Delta(h))}.
\label{chitw2}
\end{equation}
The counter term $\lambda_3$ has yet to be determined. We have the exact result,
$ \chi_t(0,h)=m(h)/h$, which from (\ref{chitw2}) implies $\tilde h +(\tilde
U+\lambda_3)m(h)$, so we can identify $\tilde
U+\lambda_3$ as $\tilde U_t(h)$ in equation (\ref{rptmf}) and the expression for
$\chi_t(\omega,h)$ with that given in equation (\ref{chitw}).
The interaction term $\tilde U_t(h)$ expressed in terms of the two
other renormalised parameters, $\tilde\Delta(h)$ and $\tilde\eta(h)$,
is
\begin{equation}
\tilde U_t(h)={h(\tilde\eta(h)-1)\over m(h)}={\pi h(\tilde\eta(h)-1)\over
{\rm tan}^{-1}(h(\tilde\eta(h)/\tilde\Delta(h))}.
\label{rut}
\end{equation}
Detailed comparison of the RPT results for the  transverse susceptibility
 based  equations (\ref{chitw2}) (or equivalently (\ref{chitw})) and   (\ref{rut})
 with a direct NRG calculations have been given  earlier \cite{Hew06}. They are
in remarkably good agreement with the NRG results for all values of the magnetic field over a magnetic
 field range $-10T_{\rm K}<h<10T_{\rm K}$,  asymptotically exact as
 $\omega\to 0$, and satisfy the Korringa-Shiba relation.\par

We can  derive an approximation for the self-energy
$\tilde\Sigma^S_{\uparrow}(\omega,h)$ analogous to that given in equation (\ref{RPAse}),
\begin{equation}
\tilde\Sigma^S_{\uparrow}(\omega,h)=\tilde U_t^2(h)\int
\tilde G^{\rm mf}_{\downarrow}(\omega+\omega',h)\chi_t(\omega',h){d\omega'\over 2\pi i}.
\label{RPTse}
\end{equation}
Some RPT
results  based on
equations (\ref{RPTse}) and (\ref{chitw2}) for the self-energy and one-electron spectral density 
have also  been given earlier  \cite{Bau07,BHO07} and compared with the
corresponding results from
direct NRG calculation. Again the agreement between the two 
sets of results was very good over a low frequency range $-0.5T_{\rm K}
<\omega<0.5T_{\rm K}$ for $h=0$ and over a larger range for higher values of
$h$. These results demonstrate 
that it is possible to find an approximation using the RPT which will
accurately reproduce the form of the self-energy in the low frequency
regime for any value of the magnetic field. \par

\section{RPT calculations in the low field limit}

 We need to extend the RPT calculations described in the previous section  to see if, given  
with renormalised parameters for one field value,
 we can  calculate the self-energy
for another field value, and hence deduce how the renormalised
parameters change as we  vary the magnetic field.
Before we consider this problem in detail 
 we consider the simpler case of using the
renormalised parameters for $h=0$ to calculate the renormalised self-energy
in the presence of a weak magnetic field $h$. 
 As we want to keep the magnetic field
term explicitly in the Hamiltonian, and as the self-energy has the form
$\Sigma_\sigma(\omega+\sigma h,h)$, we modify our procedure and use a
slightly different form for the remainder term $\Sigma^{\rm rem}_\sigma(\omega+\sigma h,h)$ via
\begin{eqnarray}&&\Sigma_\sigma(\omega+\sigma h,h)=\Sigma_\sigma(0,0) +\label{new1}\\
&&(\omega+\sigma h) {\partial \Sigma_\sigma(\omega+\sigma h,h)\over\partial\omega}\Big|_{\omega=h=0}+
   \Sigma^{\rm rem}_\sigma(\omega+\sigma h,h).\nonumber
\end{eqnarray}
 As the  renormalised parameters
are those defined for $h=0$, the renormalised self-energy and 4-vertex satisfies the equations
in (\ref{rc1}) and   (\ref{rc2}) with  $h$ set equal to zero. Substituting this
form into the equation (\ref{gf}) for the impurity Green's functions gives
\begin{equation}
G_{\sigma}(\omega,h)={z\over
    \omega-\tilde\epsilon_{\mathrm{d}}+\sigma h+i\tilde\Delta-\tilde\Sigma^r_\sigma(\omega,h)},
\label{ngf7} 
\end{equation}
so the renormalised parameters now correspond to $h=0$, and the renormalised
 self-energy is defined by
\begin{eqnarray}
\tilde\Sigma^r(\omega,h)&&=z\Big\{ \Sigma_\sigma(\omega+\sigma h,h)
- \Sigma_\sigma(0,0) -\nonumber\\
&&(\omega+\sigma h) {\partial \Sigma_\sigma(\omega+\sigma
 h,h)\over\partial\omega}\Big|_{\omega=h=0}\Big\}.
\label{new2}
\end{eqnarray}
Note that this self-energy for $h\neq 0$ is different from the 
self-energy $\tilde\Sigma(\omega,h)$  defined earlier in equation (\ref{rse}),
due to the different remainder term,
so we use a slightly different notation to distinguish them. The renormalised
self-energy $\tilde\Sigma^r(\omega,h)$ 
is required to satisfy the equation in (\ref{rc1}) for $h=0$.\par
We now show that exact result for $\tilde\Sigma^r(0,h)$ to first order in
$h$ 
  is given by  the self-consistent evaluation of the  tadpole diagram as in mean field theory.
The counter term $\lambda_1$ only cancels off the tadpole diagram for
$h=0$, so taking this into account, we get for the renormalised self-energy,
 $\tilde\Sigma^r_\uparrow(\omega,h)=-(\tilde U+\lambda_3)m(h)$, with
\begin{equation}
m(h)={1\over\pi}{\rm
  tan}^{-1}\left({h-\tilde\Sigma^r_\uparrow(0,h)\over\tilde\Delta}\right).
\end{equation}
 This equation for $m(h)$ corresponds to the exact form given by the Friedel
 sum rule. Solving these mean field equations to first order in $h$, we find 
for the zero field susceptibility,
\begin{equation}
\chi_l={\pi\tilde\Delta\over 2( 1-(\tilde U+\lambda_3)/\pi\tilde\Delta)}.
\end{equation}
This result for $\chi_l$ corresponds to the exact result given in equation (\ref{chil})
if $\tilde U+\lambda_3=\tilde U_s$. For $h=0$ we also have  $\tilde U_t=\tilde
U_s$ so this mean field equation corresponds
to the exact one given earlier in equation (\ref{rptmf2}),
but with the renormalised parameters $\tilde\Delta(h)$ and $\tilde U_t(h)$
taken at zero field. As the corrections to the zero field values for $\tilde\Delta(h)$ and
$\tilde U_t(h)$ are of order $h^2$, it follows that equation (\ref{rptmf2})
with the zero field renormalised parameters is exact to first order in
$h$. \par
\begin{figure}[!htbp]
   \begin{center}
     \includegraphics[width=0.45\textwidth]{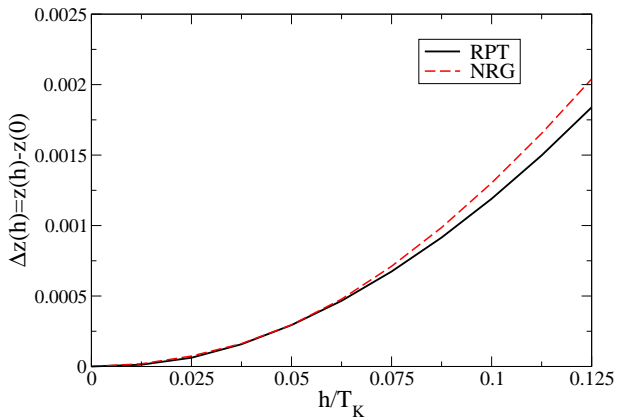}
 \caption{A plot of the change in $z(h)$, $\Delta z(h)=z(h)-z(0)$,
as a function of $h/T_{\rm K}$ calculated using the RPT for $U/\pi\Delta=3$,
 $\pi\Delta=0.1$, with renormalised prameters at $h=0$ deduced from the NRG,
compared with $\Delta z(h)$ deduced from a direct NRG calculation of the self-energy.}
     \label{dz}
   \end{center}
 \end{figure}

 
We can extend the calculations to include an $\omega$-dependence in
$\tilde\Sigma^r_\sigma(\omega,h)$ by using the RPT equation 
(\ref{RPTse}) with the zero field parameters in the transverse
dynamic susceptibility calculated from the NRG. From the change in the linear
$\omega$-dependence of the renormalised self-energy with the magnetic field
$h$, we can calculate the change $\Delta z(h)=z(h)-z(0)$. We can then compare
the results with the corresponding value of  $\Delta z(h)$ deduced from a
direct NRG calculation of the self-energy. A comparison of the two sets of
results is shown in figure \ref{dz} for the case  $U/\pi\Delta=3$,
 $\pi\Delta=0.1$. It can be seen that the two sets of results are in complete
 agreement in the very low field regime. This implies that, given the
 renormalised
parameters at $h=0$, we should be able to calculate accurately both the change
in 
$z(h)$ and the shift in the real part of the self-energy with $h$, provided we
use
a sufficiently small value of $h$. This is all the information we require to
calculate the renormalised parameters for small $h$ from the given values at $h=0$.
 \vspace*{0.7cm}

\section{Scaling equations}

We now consider how to set up a scaling equation to deduce
the renormalised parameters for a  magnetic field $h-\delta
h$, given  a renormalised self-energy that has been calculated  using the
renormalised
parameters, $\tilde\eta(h)$, $\tilde\Delta(h)$ and $\tilde U(h)$ for a
magnetic field value $h$. We first of all introduce a renormalised self-energy
$\tilde\Sigma^r_\sigma(\omega,h,\delta h)$ following the steps from equation
(\ref{new1}) to (\ref{new2}) but retain only the $\delta h$ term explicitly in
the Green's function and the remaining part is absorbed to give the
renormalised
parameters $\tilde\eta(h)$ and $\tilde\Delta(h)$ as defined earlier,
\begin{eqnarray}
\tilde\Sigma^r(\omega,h,\delta h)&&=z(h)\Big\{ \Sigma_\sigma(\omega,h-\delta h)
- \Sigma_\sigma(0,h) -\nonumber\\
&&(\omega-\sigma \delta h) {\partial \Sigma_\sigma(\omega,h)\over\partial\omega}\Big|_{\omega=\delta h=0}\Big\}.
\end{eqnarray}
The self-energy  $\tilde\Sigma^r(\omega,h,\delta h)$ satisfies the
renormalisation conditions in (\ref{rc1}) for $\delta h=0$. 
The local Green's function for the system with a magnetic field $h-\delta h$,
using the parameters for the system in a field $h$, takes the form,
\begin{eqnarray}
&&G_{\sigma}(\omega,h-\delta h)=\label{rgfh1}\\
&&{z(h)\over
    \omega-\sigma\delta h +\sigma\tilde h(h)+i\tilde\Delta(h)-\tilde\Sigma^r_\sigma(\omega,h,\delta
    h)},\nonumber
\end{eqnarray}
where $\tilde h(h)=h\tilde\eta(h)$.
However, in terms of the renormalised parameters for the field value
$h-\delta h$, this Green's function is
\begin{eqnarray}
&&G_{\sigma}(\omega,h-\delta h)=\label{rgfh2}\\
&&{z(h-\delta h)\over
    \omega+\sigma\tilde h(h-\delta h)
    +i\tilde\Delta(h-\delta h)-\tilde\Sigma_\sigma(\omega,h-\delta
    h)}.\nonumber
\end{eqnarray}
We can find an expression for the renormalised parameters $\tilde h(h-\delta
h)$ and $z(h-\delta h)$ in terms of the parameters for $\delta h=0$ and
the self-energy $\tilde\Sigma^r_\sigma(\omega,h,\delta h)$ by equating the
inverses of the Green's functions in equations (\ref{rgfh1}) and
(\ref{rgfh2}). Differentiating these with respect to $\omega$, and putting $\omega=0$, gives
\begin{equation}
z(h-\delta h)=\bar z(h,\delta h)z(h),
\end{equation}
where $\bar z(h,\delta h)=1/(1-\tilde\Sigma^{r'}_{\sigma}(0,h,\delta
h))$. Hence we get a relation between $\tilde\Delta(h-\delta h)$ and 
 $\tilde\Delta(h)$,
\begin{equation}
\tilde\Delta(h-\delta h)=\bar z(h,\delta h)\tilde\Delta(h).
\label{sc1}
\end{equation}
Equating the inverses of  (\ref{rgfh1}) and (\ref{rgfh2}) and putting
$\omega=0$, we find  a relation between $\tilde h(h-\delta h)$
and $\tilde h(h)$,
\begin{equation}
\tilde h(h-\delta h)=\bar z(h,\delta h)(\tilde h(h)-\delta h -
\tilde\Sigma^r_\uparrow(0,h,\delta h)).
\label{sc2}
\end{equation}
We can find a alternative form of this equation by expanding
$\tilde\Sigma^r_\uparrow(0,h,\delta h)$ to first order in $\delta h$ and then
using the relation, 
\begin{equation}
{\partial\tilde\Sigma^r_\uparrow(0,h,\delta h))\over
\partial\delta h}\Big|_{\delta h=0}=-\tilde\rho(0,h)\tilde U(h),
\end{equation}
which follows from the Ward identity \cite{Yam75a,Yam75b,Hew01}. We then get an equation for $\tilde
h(h-\delta h)$ in the form,
\begin{equation}
\tilde h(h-\delta h)=\bar z(h,\delta h)(\tilde h(h)-\delta h(1+R(h)))+{\rm
  O}[(\delta h)^2],
\label{sc3}
\end{equation}
where $R(h)$ is the Wilson ratio given by $R(h)=1+\tilde\rho(0,h)\tilde U(h)$.
\par
We also need to be able to calculate the new interaction term $\tilde
U_t(h-\delta h)$.  It can be seen from equation (\ref{rut}) that a knowledge of $\tilde\eta(h)=\tilde h(h)/h$
and $\tilde \Delta(h)$ is sufficient to determine $\tilde U_t(h)$.
Hence from the results for $\tilde \Delta(h-\delta h)$ and $(\tilde h-\delta h)$
we can deduce $\tilde U_t(h-\delta h)$.\par
Finally, the full 
renormalised vertex  $\tilde U(h-\delta h)$ can be calculated from the results
for  $\tilde \Delta(h-\delta h)$ and $\tilde\eta( h-\delta h)$,
by
differentiating
the magnetisation, as given in equation (\ref{magqp}), 
 to determine the static
longitudinal susceptibility, and then equating it to the expression given in
equation (\ref{chil}).

 \begin{figure}[!htbp]
   \begin{center}
     \includegraphics[width=0.45\textwidth]{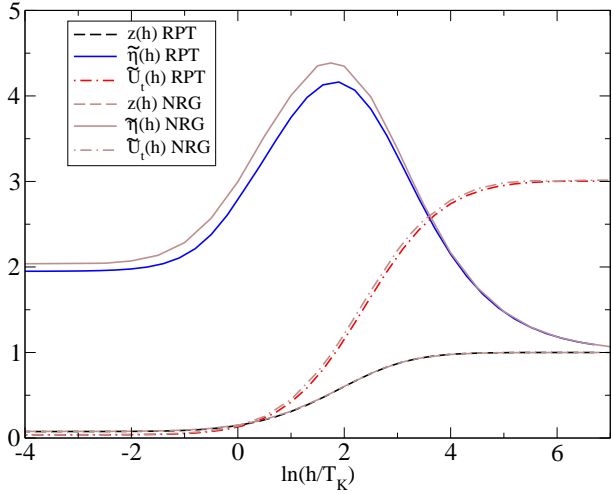}
 \caption{A comparison  the renormalised parameters,
  $z(h)=\tilde\Delta(h)/\Delta$, 
  $\tilde U_t(h)/U$ and  $\tilde\eta(h)$ as calculated from the RPT and
NRG for 
 the symmetric Anderson model, with $\pi\Delta=0.1$, $U/\pi\Delta=3$  as a function of the logarithm of the magnetic
 field $h/T_{\rm K}$, where $T_{\rm K}=\pi\tilde\Delta(0)/4$.}
     \label{rp2}
   \end{center}
 \end{figure}

We  now  have to amend the procedure outlined in the previous section to calculate
the self-energy in the reduced field $\tilde\Sigma^r_\uparrow(\omega,h,\delta h)$ so that we can
exploit
equations (\ref{sc1}) and (\ref{sc2}) or (\ref{sc3}) to extend the calculation of the renormalised
parameters to smaller values of the field.
\section{Extension to lower magnetic field values}
The Green's function for a system with a magnetic field $h-\delta h$, in
terms of renormalised parameters for a field value $h$, takes a form
 similar
to that given in equation (\ref{ngf3}) when we include the counter terms $\lambda_{1,\sigma}(h)$ and
$\lambda_{2,\sigma}(h)$, 
\begin{eqnarray}
&&G_{\sigma}(\omega,h-\delta h)=\label{nngf}\\
&&{z(h)\over
    \omega+\sigma(\tilde h-\delta
    h)+i\tilde\Delta(h)+\lambda_{1,\sigma}+\omega\lambda_{2,\sigma}-\bar\Sigma^{r}_\sigma(\omega,h,\delta h)}.\nonumber
\end{eqnarray}
The self-energy $\bar\Sigma^{r}_\sigma(\omega,h,\delta
    h)$ is now calculated with the quasiparticle propagator,
\begin{eqnarray}
&&G_{\sigma}(\omega,h-\delta h)=\label{nqpgf}\\
&&{1\over
    \omega+\sigma(\tilde h-\delta
    h)+i\tilde\Delta(h)+\lambda_{1,\sigma}+\omega\lambda_{2,\sigma}-\bar\Sigma^{r}_\sigma(\omega,h,\delta
    h)}.\nonumber
\end{eqnarray}
 The counter terms
$\lambda_{1,\sigma}$ and $\lambda_{2,\sigma}$, however, are still determined by the conditions
given in equations (\ref{rc3}) and (\ref{rc4}) but with $\delta h=0$.\par
We consider first of all the tadpole diagram, which now gives a finite
contribution because it is not cancelled completely by the counter term
for $\delta h\ne 0$. This diagram, when the cancellation due to the counter
term $\lambda_{1,\sigma}$ has been taken into account gives
$\tilde\Sigma^r_\sigma(\omega,h,\delta h)=-\sigma \tilde U_t(h)m(h,\delta h)$
where 
\begin{equation}
m(h,\delta h)={1\over\pi}{\rm
  tan}^{-1}\left({-\delta h-\tilde\Sigma^r_\uparrow(0,h,\delta h)\over\tilde\Delta}\right).
\end{equation}

  \vspace*{0.7cm}
 \begin{figure}[!htbp]
   \begin{center}
     \includegraphics[width=0.45\textwidth]{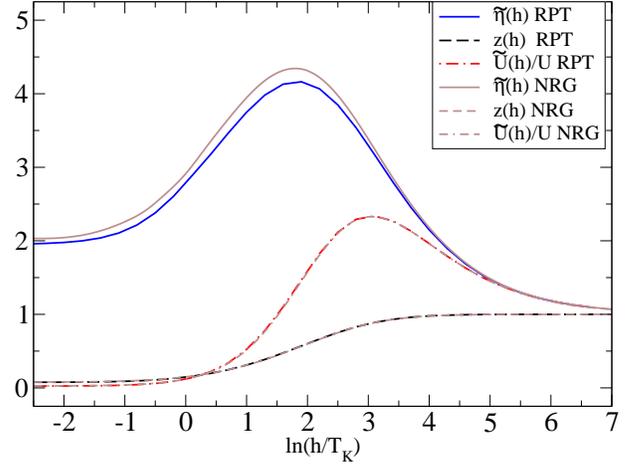}
 \caption{A comparison of the renormalised parameters,
  $\tilde\Delta(h)/\Delta$, 
  $\tilde U(h)/U$ and  $\tilde\eta(h)$ calculated using the RPT as a function of the logarithm of the magnetic
 field $h/T_{\rm K}$, compared with
  the corresponding results from the NRG given in figure \ref{rp1}.}
     \label{rp3}
   \end{center}
 \end{figure}
 \begin{figure}[!htbp]
   \begin{center}
     \includegraphics[width=0.45\textwidth]{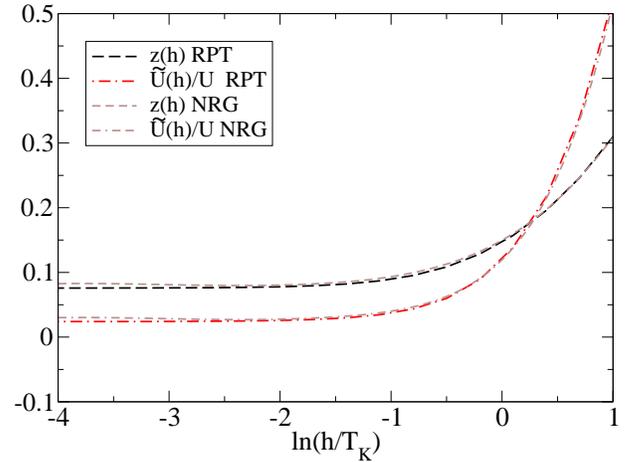}
 \caption{The NRG and RPT results for  $\tilde\Delta(h)/\Delta$ and
  $\tilde U(h)/U$ as in figure
  \ref{rp2}  over the the low field regime.  }
     \label{rp4}
   \end{center}
 \end{figure}
 \begin{figure}[!htbp]
   \begin{center}
     \includegraphics[width=0.45\textwidth]{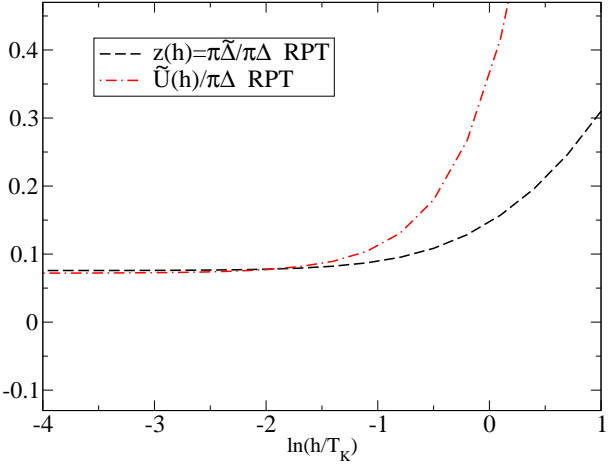}
 \caption{The renormalised parameters,
  $z(h)=\tilde\Delta(h)/\Delta$, 
  $\tilde U(h)/\pi\Delta$ as calculated from the RPT  for the same parameter
set as in figure \ref{rp1}, showing to a good  approximation
the approach to the zero field strong coupling result $\tilde U=\pi\tilde\Delta$. 
}
     \label{rp5}
   \end{center}
 \end{figure}
The mean field self-consistent equation for $m(h,\delta h)$ is then solved,
and this term absorbed into the propagator in equation (\ref{nqpgf}) so the
perturbation expansion is now about this mean field solution.
The next step is to take the repeated quasiparticle scattering term into
account in the calculation of the self-energy
$\tilde\Sigma^r_\sigma(\omega,h,\delta h)$ as in equation (\ref{RPTse}).
This constitutes our approximation for calculating
$\tilde\Sigma^r_\sigma(\omega,h,\delta h)$. The renormalised parameters 
as a function of magnetic field can now be calculated  for a field $h-\delta h$
using this approximation for $\tilde\Sigma^r_\sigma(\omega,h,\delta h)$
together with the scaling equations (\ref{sc1}) and  (\ref{sc2}) or (\ref{sc3}).\par
In figure \ref{rp2} we compare the results for $\tilde\eta(h)$, $\tilde U_t(h)$ and 
$z(h)=\tilde\Delta(h)/\Delta$ obtained in the RPT, using the scaling
equation to extend to the low field regime, with the NRG results. The results are for the symmetric model
with $U/\pi\Delta=3$, $\pi\Delta=0.1$. It can be seen that the agreement
with the NRG results is very good over the whole magnetic field range,
especially the results for  $\tilde U_t(h)$ and 
$z(h)=\tilde\Delta(h)/\Delta$. The results for  $\tilde\eta(h)$ are
smaller than the NRG results at lower field values but the difference is relatively small. In the very low field regime  $h\to 0$, the NRG results approach 
the value 2 corresponding to the Wilson ratio in the Kondo limit. The RPT
results in this regime are  smaller by approximately 3\%.\par

 \begin{figure}[!htbp]
   \begin{center}
     \includegraphics[width=0.45\textwidth]{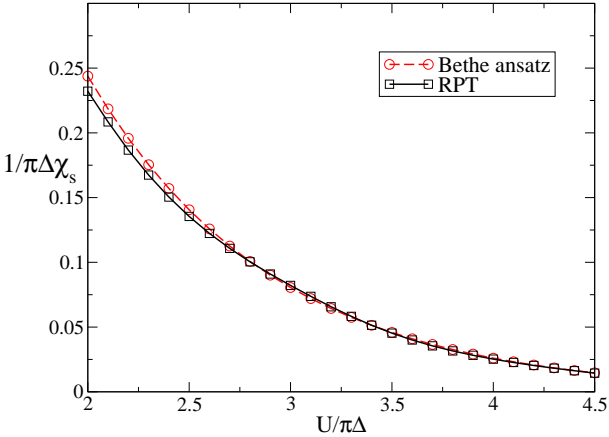}
 \caption{A comparison of $1/\pi\Delta\chi_s$ for $h=0$ as deduced from the RPT
(full curve) with the Bethe ansatz results (dashed curve) for a range of
 values of
 $ U/\pi\Delta$.  In the strong correlation regime
 $1/\pi\Delta\chi_s\to 4T_{\rm K}/\pi\Delta$.
}
     \label{Tk}
   \end{center}
 \end{figure}
 
Within the approximation we have used we can calculate $\tilde U(h)$, given 
$\tilde U_t(h)$ and $\tilde\eta(h)$, rather more simply than the method
described earlier,
 by summing the renormalised
 equivalent of the diagrams in figure \ref{gammarpa} to give
 \begin{equation}
 \tilde U(h)={\tilde U_t(h)\over 1-\tilde U_t(h)\Pi_t(0,\tilde h,\tilde\Delta(h))}.
\end{equation}
Using the fact that $\Pi_t(0,\tilde h,\tilde\Delta(h))=m_0(\tilde h,\tilde\Delta(h))/\tilde
h$, we find $\tilde U(h)=\tilde h \tilde U_t(h)/h=\tilde\eta(h)\tilde U_t(h)$. 
\par

In figure \ref{rp3} we compare the results for $\tilde\eta(h)$, $\tilde U(h)$ and 
$z(h)=\tilde\Delta(h)/\Delta$ obtained in the RPT using the scaling
 with the NRG results as given in figure \ref{rp1} for the same parameter
set. There is excellent agreement between the $\tilde U(h)$ from the RPT with the
NRG result. As  the values of $z(h)=\tilde\Delta(h)/\Delta$ and  $\tilde U(h)$
are rather small in the strong coupling regime $h\to 0$, we give an enlarged picture of the comparison
of the results in the weak field regime in figure \ref{rp4}. The agreement of the  RPT with the NRG results can be seen to be maintained down to values of the
magnetic field several orders of magnitude less than the Kondo temperature $T_{\rm K}$. Finally in figure \ref{rp5} we give a plot of the RPT values of
$\tilde U(h)/\pi\Delta$ and $\tilde\Delta(h)/\Delta$. The fact that the two
curves merge as $h\to 0$ shows that the RPT results asymptotically  satisfy the relation
$\tilde U(0)=\pi\tilde\Delta(0)$, as given in equation (\ref{KT}) corresponding to a
single renormalised energy scale.\par 
In figure \ref{Tk} we compare the results for $1/\pi\Delta\chi_s$  from the RPT
calculation in the zero field limit for a range of values of $U/\pi\Delta$
with the corresponding Bethe ansatz results \cite{HZ85} given by
\begin{equation}
{1\over \pi\Delta \chi_s}= {4T_{\rm K}\over (1+I)\pi\Delta},\end{equation}
where 
\begin{equation}
{T_{\rm K}}=\left({U\Delta\over 2}\right)^{1/2}e^{-\pi
  U/8\Delta+\pi\Delta/2U},\end{equation}
and
\begin{equation}
I={1\over \sqrt{\pi}}\int_0^{\pi\Delta/2U} {e^{x-\pi^2/16x}\over\sqrt{x}}\,
  dx.
\end{equation}
For $U/\pi\Delta>2$, the integral term $I$ is very small compared to unity
so that in this regime ${1/\pi\Delta \chi_s}$ corresponds to ${4T_{\rm
    K}/\pi\Delta}$. \par
It can be seen from this comparison there is very good agreement with the
Bethe ansatz results in the strong correlation regime $U/\pi\Delta>2.5$.
It shows clearly that the RPT results give the correct form for 
$T_{\rm K}$ in the Kondo regime, not only in the exponential dependence on $U$
but also in the prefactor. There is a progressive improvement in the agreement with the
exact Bethe ansatz results with increase of $U/\pi\Delta$ over the range $2\le U/\pi\Delta\le 4.5$.\par

 It might seem surprising to be able to get such 
precise agreement by taking what would appear to be only a subclass
of diagrams into account. The explanation is that we are considering the low
energy regime and, for strong correlation, the dominant low energy scattering 
is with  the spin fluctuations. The low energy charge fluctuations
are suppressed in this regime and their effects can be taken into account
by suitably renormalised vertices. For smaller values of $U/\pi\Delta$ the
charge fluctuations begin to play a role and have to be
considered  explicitly. In this regime they  cannot be taken into account simply  
by the  use of an effective frequency independent vertices. \par

\vspace*{0.7cm}
 \begin{figure}[!htbp]
   \begin{center}
     \includegraphics[width=0.45\textwidth]{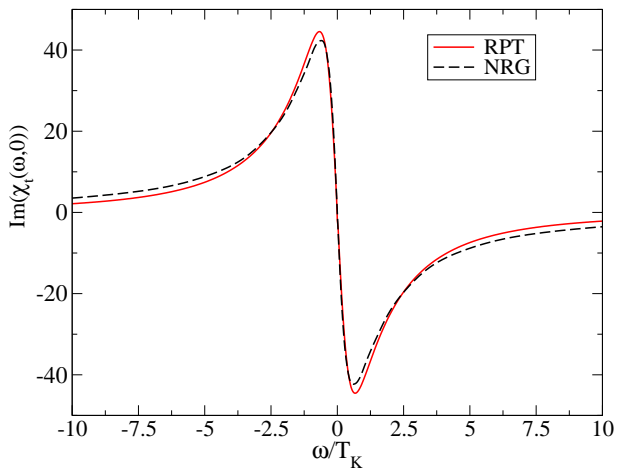}
 \caption{A plot of ${\rm Im}\chi_t(\omega,0)$
versus $\omega/T_{\rm K}$ for $U/\pi\Delta=3$, $\pi\Delta=0.1$ and $h=0$,
  as calculated directly from the NRG (dashed curve) and
from the RPT with renormalised parameters also calculated using the RPT.}
     \label{ssrpt0}
   \end{center}
 \end{figure}
 \noindent
 \begin{figure}[!htbp]
   \begin{center}
     \includegraphics[width=0.45\textwidth]{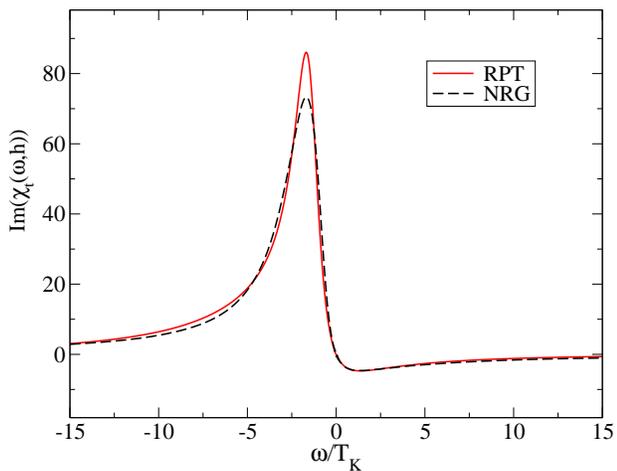}
 \caption{A plot of ${\rm Im}\chi_t(\omega,h)$
versus $\omega/T_{\rm K}$ for $U/\pi\Delta=3$, $\pi\Delta=0.1$ and $h=T_{\rm K}=0.002$,
  as calculated directly from the NRG (dashed curve) and
from the RPT with renormalised parameters also calculated using the RPT.}
     \label{ssrpth}
   \end{center}
 \end{figure}

We are now in a position to calculate the low energy dynamic susceptibility
and self-energy at $h=0$ and low fields {\it entirely} from the 
RPT. 
In figures \ref{ssrpt0} and \ref{ssrpth} we compare the RPT results for the imaginary
part of the dynamic transverse susceptibility (full curve) with the corresponding results
 calculated directly from the NRG (dashed curve) for the case  $U/\pi\Delta=3$ and $\pi\Delta=0.1$
and $h=0$ and $h=T_{\rm K}$ respectively. These have been calculated using equation (\ref{chitw})
for $\chi_t(\omega,h)$  with the renormalised
parameters. 
It can be seen that they are in good agreement
over the whole low frequency regime. The discrepancy in the peak height in figure  \ref{ssrpth}
is almost certainly due to the logarithmic broadening used in the NRG calculations which has the effect of
reducing  the height of any peak displaced from the origin. The higher the magnetic field value the larger
the effect becomes as the peak gets shifted further from the origin \cite{Hew06}.\par
\begin{figure}[!htbp]
   \begin{center}
     \includegraphics[width=0.45\textwidth]{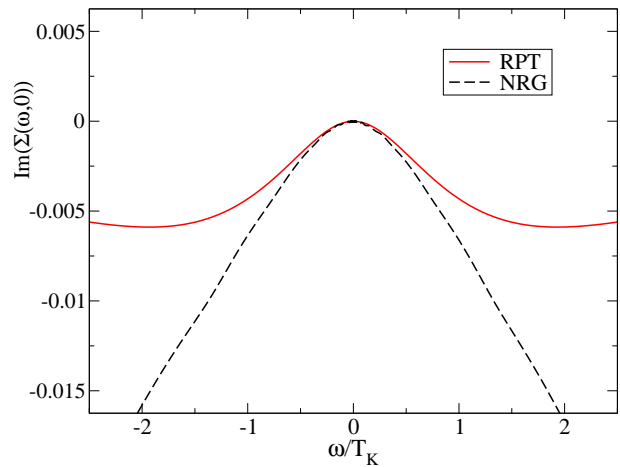}
 \caption{A plot of the imaginary part of the self-energy $\Sigma_\uparrow(\omega,0)$
versus $\omega/T_{\rm K}$ for $U/\pi\Delta=3$, $\pi\Delta=0.1$ and $h=0$,
  as calculated directly from the NRG (dashed curve) and
from the RPT with renormalised parameters also calculated using the RPT.}
     \label{serpt0}
   \end{center}
 \end{figure}
\begin{figure}[!htbp]
   \begin{center}
     \includegraphics[width=0.45\textwidth]{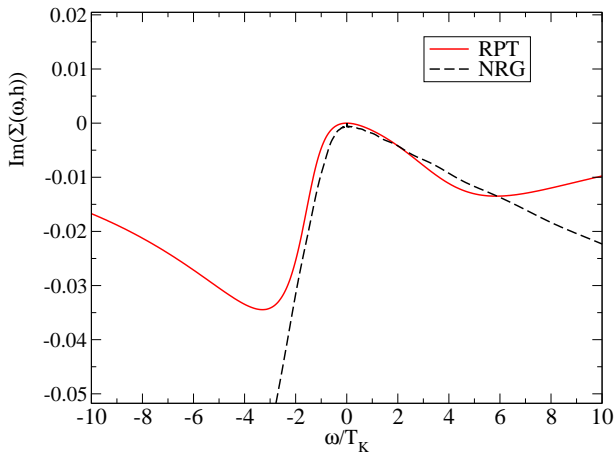}
 \caption{A plot of the imaginary part of the self-energy $\Sigma_\uparrow(\omega,h)$
versus $\omega/T_{\rm K}$ for $U/\pi\Delta=3$, $\pi\Delta=0.1$ and $h=T_{\rm K}$,
  as calculated directly from the NRG (dashed curve) and
from the RPT with renormalised parameters also calculated using the RPT.}
     \label{serpth}
   \end{center}
 \end{figure}

 In  figures
\ref{serpt0} and \ref{serpth} we make a similar comparison of the RPT results for the imaginary part  of the
self-energy calculated to second order in $\tilde U$
 for the same set of parameters, $U/\pi\Delta=3$ and $\pi\Delta=0.1$
and $h=0$ and $h=T_{\rm K}$. 
 The imaginary part of the  self-energy $\Sigma_{\sigma}(\omega,h)$
 is related to the imaginary part of the renormalised self-energy
 $\tilde\Sigma_{\sigma}(\omega,h)$ from equation (\ref{rse})
 via
 \begin{equation}
 {\rm Im}\Sigma_{\sigma}(\omega,h)={1\over z(h)}\,{\rm
   Im}\tilde\Sigma_{\sigma}(\omega,h).
 \end{equation}
 There is very good agreement between the two sets of curves over the range $|\omega|\le 0.5T_{\rm K}$.
The small discrepancy in the low frequency regime in figure \ref{serpth} could be due to the imaginary
part of the self-energy in the NRG results does not precisely equal zero at
$\omega=0$. Surprisingly, using the RPT only to second order in $\tilde U$ for $h=0$ extends the range
of the agreement to $|\omega|\le T_{\rm K}$. \par
 The low order RPT calculations give the 
asymptotically exact results corresponding to Fermi liquid theory in the low
frequency regime. However, once the renormalised parameters have been calculated the renormalised
perturbation theory is completely defined, and can be extended to higher
frequency and higher energy scales by taking higher order diagrams into
account. The extension to the higher frequency range is currently being
studied, and some preliminary results have been published \cite{Hew01,BHO07}.

 \noindent
\vspace*{0.7cm}

\section{Conclusions}
We have set out to see whether we can access the low energy behaviour of a
strongly correlated system using a renormalised form of perturbation theory
which would be applicable to a general class of models. We have taken as a
test case the particle-hole symmetric single impurity Anderson model with
parameters in the strongly correlated  Kondo regime, where we
have exact results  which can be used to test any approximation  used. The
approach has been based on the renormalised perturbation theory, which has
previously  been
shown to give asymptotically exact results for this model in the low energy
regime in terms of renormalised parameters. Hitherto  these
renormalised parameters have been calculated in terms of the bare parameters that specify the
model from an analysis of the low energy fixed point of an
NRG calculation. The renormalised parameters can be
determined very accurately using this indirect approach, but this makes the method
dependent on the NRG.
 Our goal has been to find an
alternative and more general way to calculate these parameters, and to test
the approximations used by comparing the results with those derived using the NRG.
 To do this we have exploited the fact that the
low energy spin fluctuations, which cause the strong renormalisation effects,
can be suppressed by the application of a very strong magnetic field,
allowing standard perturbation theory to be applied and the renormalised parameters to
be calculated in the large field regime. The renormalised perturbation theory
is then used to calculate the renormalised parameters on reducing the field
$h$ by a small amount to $h-\delta h$, so setting up a scaling relation.
The solution of this scaling equation allows the field value to be extended
down to $h=0$. The approximation used has
been based on a mean field-like theory at each stage with a self-energy that
includes the RPA-like fluctuations about the mean field. The results for the
renormalised parameters have been compared with those calculated in
previous work using the NRG for the complete magnetic field range. This
relatively simple approximation scheme gives  remarkably good results for
magnetic field values 
down to $h=0$, as demonstrated in the comparison with the Bethe ansatz results
for the zero temperature susceptibility.   This calculation demonstrates that it is possible to access the strong
correlation regime using a perturbational analysis. \par
There are some similarities with the approach used here for the Anderson model
 with two other perturbational
approaches.  In the local moment approach \cite{LET98,LD01b} RPA diagrams are
included in a two self-energy formalism, where the self-energies are
constructed from a  mean field broken symmetry state for $h=0$. The symmetry, which should be
restored by the low energy dynamics, is then imposed by the
 requirement that the imaginary part of self-energy
is zero at $\omega=0$, which determines the mean field magnetisation. This approach has an advantage in giving
an interpolation such that the high energy features, corresponding to the
atomic
states, are included, and does give a low energy scale $T_{\rm K}$ which
depends exponentially on $U/\pi\Delta$ with an exponent in agreement with the
Bethe ansatz result \cite{TW83}. The fact that the symmetry has to be imposed, however, rather
than develop naturally as the field is reduced in the RPT approach, is a
limitation.\par
The other related approaches have been  based on the functional renormalisation group.
  A recent example is the work of Bartosch et al. \cite{BFRK09} who use a 
Hubbard-Stratonovich transformation and include  both longitudinal and transverse
spin fluctuations in calculating the self-energy. A cut off
$\Lambda$ was
imposed to suppress the low energy
 spin fluctuations, which plays a role similar to the large  applied field
used in the RPT.
 Flow equations  for the renormalised vertices
 as a function of $\Lambda$  were then derived based using various approximations based on the
truncation of the higher order vertices.
The results, however, for the quasiparticle
weight factor $z$, which should behave as  $z\sim T_{\rm K}/\Delta$ in the
Kondo regime, did not have the  exponential dependence on $U/\pi\Delta$. \par

The demonstration of the feasibility of this RPT approach has been for a specific
model, for a particular parameter set. The method, however, is a rather
general
one, and should be applicable to more general  impurity models. A good test
case would be the general
$n$-channel Anderson model with a Hund's rule exchange term. The
renormalised parameters have been calculated for this model using the NRG for $n=2$
\cite{NCH10}, but  for $n>2$ the NRG calculations  become prohibitively difficult
due to the sizes of the matrices to be diagonalised.  
It should also be possible to apply the approach to lattice models of strongly
correlated systems, where  in an
appropriate applied field the low energy behaviour of the system
evolves continuously on reducing an appropriate applied  field to zero.  
 The next step is to test
the approach for these  more general types of
models. \par

 \noindent

\bigskip\par
\noindent{\bf Acknowledgment}\par
\bigskip
{We thank Johannes Bauer, Akira Oguri, Yunori Nishikawa and Daniel Crow  for helpful
  discussions, and EH acknowledges the support of an EPSRC grant.}

\bibliography{artikel}
\bibliographystyle{h-physrev3}

\end{document}